\documentclass[reprint,twocolumn,secnumarabic,amssymb, nobibnotes, aps, pre,showkeys]{revtex4-2}
\usepackage{bm,chemformula,graphicx,titlesec}
\usepackage[unicode]{hyperref}
\hypersetup{colorlinks= true,
	citecolor= blue,
	linkcolor= red,
	urlcolor=magenta,
	filecolor=cyan,
	linkbordercolor={1 0 0},
	citebordercolor={0 1 0},
	urlbordercolor={0 1 1},
} 
\usepackage[mathlines]{lineno}
\usepackage[section]{placeins}
\usepackage{tikz}
\usepackage{float,wrapfig}
\usepackage{microtype}
\usepackage[tight]{subfigure}
\begin{document}
	\preprint{APS/123-QED}
	\title{Regulating protocols of globally coupled continuum systems: Effects on dynamics and thermodynamics}
	\author{Premashis Kumar}
	\email{pkmanager@bose.res.in}
	\affiliation{S. N. Bose National Centre For Basic Sciences, Block-JD, Sector-III, Salt Lake, Kolkata 700 106, India}
	\date{June 6, 2024}
	\begin{abstract}
		Controlling and understanding phenomena in coupled systems remains a significant challenge across diverse fields. This study investigates a simple globally coupled chemical system that exhibits a range of rich collective dynamics, from coherence to chimera states, controllable by a system parameter. We explore the effects of altering this control parameter using two different protocols. In protocol-I, a continuous variation of the control parameter leads to memory effects, resulting in two distinct sets of phenomena in the forward and reverse directions. In contrast, protocol-II, which includes a resetting feature, yields an entirely distinct collection of behaviors. Additionally, we capture the evolution of key elements of nonequilibrium thermodynamics associated with these dynamical states under both protocols, revealing distinct signatures for each. This work highlights the complexity and uniqueness of regulating coupled systems via control parameters compared to corresponding single systems, and it carries potential implications for the manipulation and application of collective behaviors.              		
	\end{abstract}
	\maketitle
	\section{\label{intro}Introduction}
	A coupled nonlinear system has been described and employed in various forms with diverse coupling schemes~\cite{Aronson1990AmplitudeRO, PhysRevLett.80.5109, conjucup, KOSESKA2013173} and ingredients to contemplate diverse real-world and empirical collective phenomena. Specifically, populations of coupled oscillators~\cite{WINFREE196715, Kuramoto1984ChemicalTurbulence} have been used as simplified models to capture and explain essential behaviors in biological and chemical systems, such as rhythmicity in pancreatic and cardiac cells~\cite{glass2001synchronization, strogatz1993coupled}, the synchronous flashing of fireflies~\cite{buck1976synchronous,mirollo1990synchronization}, the unison chirping of crickets~\cite{mirollo1990synchronization}, oscillatory patterns in neural networks~\cite{andreas11}, periodic or doubly periodic behavior in the array of Josephson junctions~\cite{swift1992averaging, Josephson02}, multistability and cluster patterns in coupled chemical systems~\cite{BARELI1985242, Epsteinoscillatorycluster}, counterintuitive chimera state~\cite{kuramoto2002coexistence, strogatz, Tinsleynature2012} and many more. Nevertheless, the existence of rich and complex emergent behaviors~\cite{epsreview, canoav} makes these investigations of coupled oscillators intriguing and challenging. Often, a complete knowledge of the system components and interaction scenarios and the substantial study of coupling schemes are inadequate for elucidating, characterizing, elaborating, and predicting these collective behaviors. Therefore, coupled systems still draw attention from different disciplines.         
	
    It is possible to extend the oscillator models by incorporating the spatial dimension. This spatially-extended mathematical model can be identified as a reaction-diffusion system (RDS)~\cite{Turing1952TheMORPHOGENESIS}, a ubiquitous mathematical framework for visualizing various real-life phenomena involving spatial and temporal dynamics~\cite{murray1, Cross2009PatternSystems, hoylebook}. In this type of model, components of the system interact with each other and can also diffuse over the spatial dimension. As a result, the system dynamics within a finite space can be captured by combining interaction terms with diffusion terms. Particularly when an RDS operates near the onset of Hopf instability~\cite{strogatz:2000}, the complex Ginzburg-Landau equation~\cite{Kuramoto1984ChemicalTurbulence, aranson2002world} (CGLE) can capture the essential behavior of a nonlinear oscillatory system. Importantly, this CGLE has a universal form~\cite{Cross2009PatternSystems}, which allows exploring dynamic features regardless of specific systems. The corresponding discrete variant of the CGLE is known as the Stuart-Landau equation, derived in the context of pattern-forming instability~\cite{stuart1960nonlinear}. However, in the investigation of collective dynamics, these heuristic equations can be modified by embracing various coupling scenarios~\cite{CGLEKURA, NakagawaKuramoto, HakimCGLE, cgleglobal, SethiaSen1, sethiasen, mcgle1st, mcgle2} to facilitate the emergence of various collective states mentioned earlier.
	
	Nevertheless, coefficients of these modified heuristic equations can be explicitly specified by parameters of a particular RDS and thus provide flexibility and insight in exploring and predicting the behaviors of coupled systems by controlling system parameters. Indeed, within a limited-resource environment, obtaining desired responses from coupled nonlinear systems concerning some externally controlled parameters is a crucial line of investigation and can reflect real-world situations properly. Till now, research related to control parameter changes in coupled systems has predominantly focused on identifying instabilities, finding emergent behaviors, and capturing transitions among various collective behaviors over specific parameter spaces. However, these studies often overlook the specific protocols regarding parameter alteration within the system. While this detail may be irrelevant in many cases, it can be crucial for understanding the dynamics of a population of coupled oscillators, which can exhibit a range of stabilities, from bistability to multistability. In this report, we aim to address this gap by emphasizing the importance of indicating proper protocols for parameter alterations in coupled systems.  
	
	This investigation is performed within the framework of the globally coupled continuum system in which coefficients of the amplitude equation are acquired from a prototypical RDS. In a similar globally coupled system, previously chimera state was realized and characterized thermodynamically at the fixed values of the system parameters and coupling parameter~\cite{pkgg3}. Then, a range of collective states was also generated around this chimera state by varying the coupling parameters~\cite{pkgg6}. However, in both of these investigations, the system parameters were kept fixed at some specific values only. We aim to expand the scope of these investigations of the coupled system and capture the effect of the control parameter change on the dynamics and thermodynamics of this coupled continuum system. Therefore, we will acquire the different collective dynamics using different protocols of altering the control parameter and then also characterize those collective states using nonequilibrium thermodynamic entities~\cite{pkgg3, pkgg6}. Studying this control parameter effect on the dynamics and thermodynamics of the same prototypical system was previously done for the single oscillator in different contexts~\cite{Falasco2018InformationPatterns,pkgg}. In those particular contexts, these two protocols for changing the control parameter are equivalent. However, we have identified that the control parameter change in the coupled system by these protocols will have different consequences. Then, it will be interesting to compare the dynamics and corresponding thermodynamic costs generated by two different protocols with the same range of control parameters, while all other system and coupling parameters are kept fixed. For acquiring the globally coupled system dynamics, we essentially implement the same ansatz as in Refs.~\onlinecite{pkgg3} and~\onlinecite{pkgg6}. Additionally, for the identification of states in this coupled system, we employ similar quantitative metrics, the strength of incoherence (SI), $S$, and the discontinuity measure (DM), $H$ as in those two references.
	
	The layout of the paper is as follows: First, we describe the globally coupled system in Sec.~\ref{sysdes}. In the next section, protocols for altering the control parameter are depicted, and corresponding concentration dynamics are properly formulated. In Sec.~\ref{ntf}, a nonequilibrium thermodynamic framework has been provided. Results and discussion are presented in Sec.~\ref{rnd}. Finally, we summarize and conclude the paper in Sec.~\ref{conu}.
    \section{System description \label{sysdes}}
	\begin{figure*}
		\centering 
		\includegraphics[width=\textwidth]{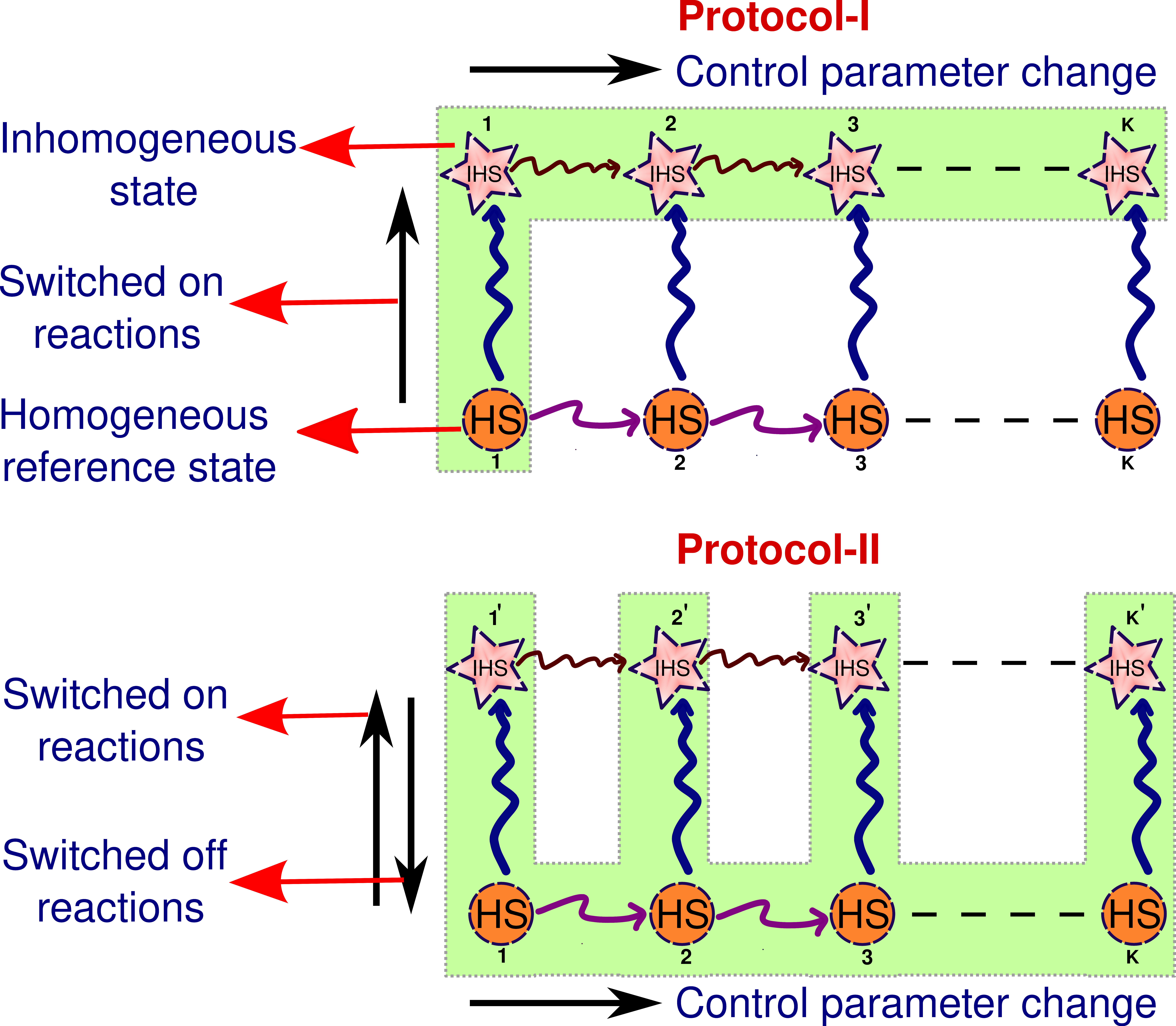}
		\caption{\label{protocol}Schematic of two different protocols: (a) For protocol-I, the system starts at a homogeneous reference state (`HS' in the figure), and after the time interval $t$, an amplitude snapshot is captured (`IHS' in the figure). The snapshot related to the previous control parameter is utilized for acquiring the dynamics at subsequent control parameter values. (b) In protocol-II, the system dynamics are reset (illustrated as switched-off reactions) after each control parameter value.}
	\end{figure*}
	For the collective dynamics investigation, we consider an abstract chemical oscillator model, the Brusselator~\cite{Prigogine1968SymmetryII, Nicolis1977Self-organizationFluctuations}. All the collective states are generated near the onset of the Hopf instability point of the single Brusselator. However, instead of explicitly incorporating coupling in the RDS description of the Brusselator, we have enacted coupling at the level of the Brusselator CGLE, as discussed in~\ref{selampli}. This approach adequately captures the essential dynamics of amplitude-mediated collective states, simplifying the dynamic and thermodynamic investigation. Due to the universality of the CGLE description, analogous collective states can also be realized in a broader globally coupled system, beyond the Brusselator. 
	\subsection{\label{model}Brusselator near Hopf instability}
	A reversible variant of Brusselator is described by the following four elementary chemical reactions, 
	\begin{equation}
		\begin{aligned}
			\rho&=1:&\ch{A&<=>[\text{k\textsubscript{1}}][\text{k\textsubscript{-1}}] X}\\
			\rho&=2: &\ch{B + X&<=>[\text{k\textsubscript{2}}][\text{k\textsubscript{-2}}]Y + D}\\
			\rho&=3:& \ch{2 X + Y&<=>[\text{k\textsubscript{3}}][\text{k\textsubscript{-3}}]3X} &\textsf{(Autocatalytic)}\\ 
			\rho&=4:&\ch{X&<=>[\text{k\textsubscript{4}}][\text{k\textsubscript{-4}}]E},
		\end{aligned}
		\label{crn}
	\end{equation}
	with $'\rho'$ being the reaction step label. Species, $\{X, Y\}\in I$, with dynamic concentration profiles, are regarded as intermediate species, and externally controllable species, $\{A, B, D, E\}\in C$ are chemostatted species with homogeneous and constant concentration. By investigating the intermediate species' dynamics concerning the chemostatted species, we often gain the proper insights into the operation of this chemical reaction network (CRN). Further, incorporating diffusion into this reaction network, the reaction-diffusion system (RDS) of the Brusselator in one spatial dimension $r\in [0,l]$ yields, 
	\begin{equation}
		\begin{aligned}
			\dot{x}&={k_1}a-({k_2}b+k_4)x+{k_3}x^2y+D_{11}x_{rr}\\
			\dot{y}&={k_2}bx-{k_3}x^2y+D_{22}y_{rr}. 
			\label{ddynamic}
		\end{aligned}
	\end{equation}
	Here $x=[X],y=[Y],b=[B],a=[A]$ denote concentrations of species and $D_{11}$ and $D_{22}$ are self-diffusion coefficients of $X$ and $Y$, respectively. In Eq.~\eqref{ddynamic}, the forward reaction rate constants $k_{\rho}$ are assumed to be significantly larger than the corresponding reverse reaction rate constants, i.e., $k_{\rho}\gg k_{-\rho}$. 
	
	In the absence of diffusion, the homogeneous steady-state of the Brusselator is, 
	$x_{0}=\frac{k_1}{k_4}a$, $y_{0}=\frac{{k_2}{k_4}}{{k_1}{k_3}}\frac{b}{a}$. However, by changing the system parameter, we can generate Hopf instability leading to homogeneous oscillation in the system dynamics. The critical control parameter value for the Hopf instability is acquired, $b_{cH}=\frac{k_{4}}{k_{2}}+\frac{k_{1}^2 k_{3}}{k_{2}{k_4}^2}a^2$ using the linear stability analysis around the steady-state values $(x_0,y_0)$. The critical frequency of the oscillation turns out, $f_{cH}=\sqrt{\frac{k_{1}^2 k_{3}}{k_4}}a$ and the critical eigenvector, $U_{cH}$, is derived as,
	$U_{cH}=
	\begin{pmatrix}
		1+\frac{i}{a}\sqrt{\frac{k_4}{k_3}}\frac{1}{k_1}\cr
		-(1+\frac{{k_4}^3}{k_3 {k_1}^2}\frac{1}{a^2}) 
	\end{pmatrix}$. Further, utilizing the Hopf instability onset condition in the linearized description of Brusselator RDS (Eq.~\eqref{ddynamic}), the critical value of the control parameter, $b$ becomes
	$b_{ctw}=\frac{k_{4}}{k_{2}}+\frac{k_{1}^2 k_{3}}{k_{2}{k_4}^2}a^2+\frac{(D_{11}+D_{22})}{k_2}q^2$, where the wave number, $q=\frac{2n\pi}{l}$, according to periodic boundary conditions in the finite domain of size $l$ with $n$ being an integer. Hopf instability conditions impose an additional restriction on the wave number selection via the following relation,
	$D_{22}^2q^4-[D_{11}-D_{22}]\frac{k_{1}^2k_3}{{k_4}^2}(aq)^2-\frac{{k_1}^2k_3}{k_4}a^2\le0$.
	\subsection{\label{selampli}Globally coupled amplitude equation related to Brusselator RDS}
	Near the Hopf bifurcation point, the CGLE can provide a faithful description and qualitative insights into Brusselator RDS dynamics. The normal form of the CGLE~\cite{Nicolis1995IntroductionScience, Cross2009PatternSystems, Walgraef1997ThePatterns} in spatially extended system is  
	\begin{equation}
		\frac{\partial Z}{\partial t}=\lambda Z -(1-i\beta)\mid Z \mid ^2Z+(1+i \alpha)\partial_{r}^2 Z,
		\label{ncgle}
	\end{equation}
	where $Z$ is the complex amplitude field. For the Brusselator RDS, we acquire coefficients, $\lambda =\frac{b-1-a^2}{2}$, $\beta=\frac{4-7a^2+4a^4}{3a(2+a^2)}$, and $\alpha=\frac{a(D_{22}-D_{11})}{(D_{11}+D_{22})}$,  and these coefficients encompass the details of the Brusselator RDS. These coefficients can be obtained by the Krylov-Bogolyubov (KB) averaging method~\cite{pkgg}.                                                                                                                                                                                                                                                                                                                                                                                                                                                                                                                                                                  
	
	In this investigation, we will explore the scenario of global coupling in a continuum chemical oscillatory system. For that purpose, we enforce this coupling at the level of amplitude dynamics of the system, instead of the RDS description. Particularly, our ansatz is that a globally coupled chemical system near the onset of Hopf instability can be effectively represented by a modified complex Ginzburg-landau equation (MCGLE) with nonlinear global coupling~\cite{pkgg3}. This MCGLE has been exploited to encapsulate a rich variety of spatiotemporal patterns, including chimeras, in the experiment of photoelectrochemical oscillators~\cite{schmidtglobal}. The coefficients of the MCGLE equation need to be quantified in terms of parameters of Brusselator RDS to realize the dynamics of a globally coupled Brusselator analytically. 
	
	In the presence of the additional nonlinear global coupling in Eq.~\eqref{ncgle}, the MCGLE~\cite{mcgle1st,mcgle2} reads, 
	\begin{eqnarray}
		\frac{\partial Z}{\partial t}=\lambda Z -(1-i\beta)\mid Z \mid ^2Z+(1+i \alpha)\partial_{r}^2 Z \nonumber \\                     
		-(\lambda+i\nu) \left\langle Z \right\rangle
		+(1-i\beta)\left\langle \mid Z\mid ^2Z \right\rangle,
		\label{gcncgle}
	\end{eqnarray}
	where $\left\langle...\right\rangle$ denotes the spatial average and the mean field oscillation is  $\left\langle Z \right\rangle=Z_0=\eta \exp(-i\nu t)$ with $\eta$ and $\nu$ being the amplitude and frequency, respectively. Further, the threshold value of $\eta$ is obtained as $\eta_c=\sqrt{\frac{\lambda}{2}}$, below which uniform oscillation becomes unstable irrespective of other parameter values. Here, the coefficient of CGLE $\lambda$ contains the Brusselator's parameters, $a$ and $b$. Thus, the coupling parameter $\eta$ is dictated by the Brusselator's parameters, emphasizing the connection between the globally coupled amplitude equation employed here and the Brusselator system. $\eta$ can be understood as the coupling strength of the system. It is worth noting that in previous descriptions of globally coupled CGLE~\cite{NakagawaKuramoto, cgleglobal}, the coupling strength was included as a multiplicated factor of the coupling term. In contrast, $\eta$ is not explicitly present in the considered MCGLE. The nonlinear global coupling described in the MCGLE can be visualized as an effective external force acting on each Brusselator.
	\section{\label{proto} Protocols for altering the control parameter and obtaining amplitude field}
	The MCGLE, i.e., Eq.~\eqref{gcncgle} has been solved numerically by exploiting a pseudospectral method incorporated with an exponential time stepping algorithm~\cite{COX2002} under two different protocols shown in Fig.~\ref{protocol}. We vary the control parameter, $b_n$, over a range of values ($5.10$ to $5.30$), keeping the other parameter, $a$,  fixed ($a=2$). Here, $n=1, 2, 3,..., K$ refers to the different values of the parameter $b$ and $K$ is the number of $b$ values considered. As illustrated in Fig.~\ref{protocol}, we will implement the following two tractable protocols for altering the control parameter, which will generate two different kinds of dynamics in this globally coupled chemical system. Homogeneous reference states (`HS' in the figure) in both protocols are denoted as $1, 2, 3,..., K$. However, to segregate the captured amplitude snapshots (`IHS' in the figure) of these two protocols, we use $\prime$, i.e., $1^\prime, 2^\prime, 3^\prime,..., K^\prime$ in the schematic of protocol-II in~\ref{protocol}. 
	\subsection{Protocol-I: a continuous variation in the control parameter values}
	In this protocol, designated as $\mathcal{CV}$, we start with a uniform state of the amplitude at a control parameter value referred to as $b_{n=1}$ and acquire an amplitude snapshot, $A_{M_1}^\mathcal{CV}$ of an inhomogeneous state from the MCGLE at a time point, $t_1<T$. This $A_{M_1}^\mathcal{CV}$ depends on the system parameters ${b_{1}, a}$ and the time point $t_1$. Above the time point $t_1$, the control parameter assumes the next value, $b_{n=2}$, and the amplitude snapshot derived at $b_{n=1}$ serves as the initial state of the MCGLE dynamics. The process continues till the time point $t_2$ when the control parameter is changed again to the next value. We repeat this step for the successive control parameter values up to time $T$ without returning to the uniform state. Since the snapshot of the MCGLE integration at a previous control parameter value, $b_{n-1}$, supplies the initial state to obtain the MCGLE amplitude for the present control parameter value, $b_{n}$, we recognize this protocol as the continuous change in the control parameter. It is important to highlight that the time counting is on throughout the control parameter variation, and amplitude data corresponding to control parameter values are obtained at fixed time intervals, $t_n-t_{n-1}=\tau$. So, in this protocol, the system spends $\tau$ time at a particular control parameter value before the subsequent alteration of the parameter. Following this protocol, we continue the process until time, $T$, initiating at time, $0$. Additionally, we assume that the chemostats (chemical reservoirs) attain equilibrium instantaneously, relative to the time scale of system dynamics.
	
	The amplitude fields acquired via protocol, $\mathcal{CV}$ are then employed to assess the spatiotemporal collective concentration dynamics of the coupled Brusselator system explicitly according to the ansatz of ref.~\cite{pkgg3, pkgg6},  
	\begin{equation}
		{z_{IH_n}}^\mathcal{CV}=z_{I0_n}+A_{M_n}^\mathcal{CV}U_{cH}\exp(i nf_{cH}\tau)+C.C..
		\label{chwave}
	\end{equation} 
	Here $A_{M_n}^\mathcal{CV}$ is the amplitude field numerically obtained from the Eq.~\eqref{gcncgle}, and  ${z_I}_{0n}$ represents the initial uniform concentration field, corresponding to $n$th value of the control parameter. $U_{cH}$  and $f_{cH}$ are the critical Eigenvector and the natural frequency, respectively, of the Brusselator within the Hopf instability regime as obtained in sec.~\ref{model}.    
		\begin{figure*}
		\centering 
		\includegraphics[width=\textwidth]{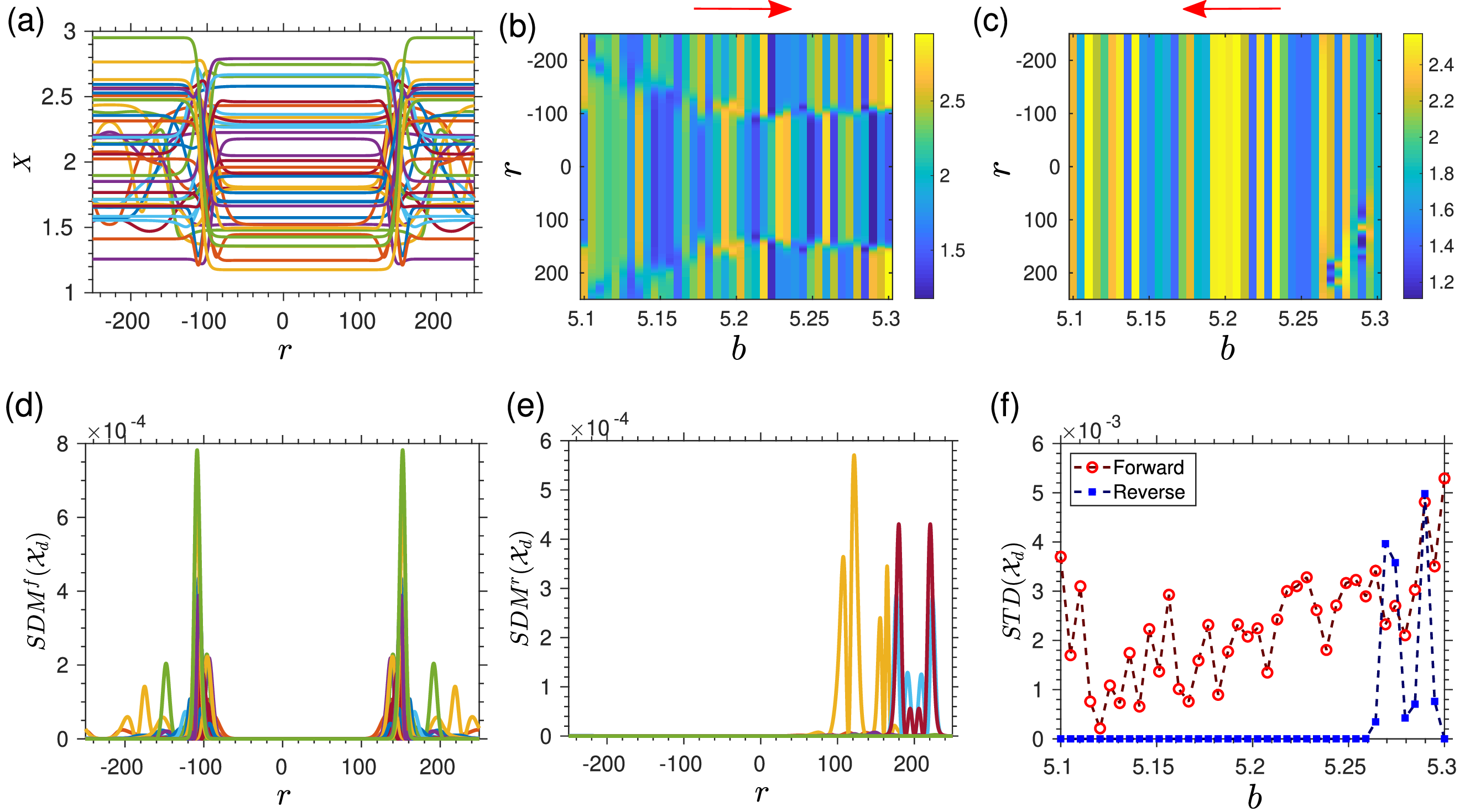}		
		\caption{\label{yaconcentration}(a) Line plots of the activator concentration, $X$, over the spatial dimension, $r$, correspond to the range of control parameter values (b) The concentration field image of the activator resulting from the forward variation of the control parameter $b$ (c) The reverse counterpart of the concentration field image of the activator is acquired. Arrows in the figures indicate the direction of the control parameter change. The squared deviations from the mean (SDM) of the entity $\mathcal{X}_d$ are illustrated for (d) the forward variation and (e) the reverse variation of $b$. The superscripts `$f$' and `$r$' denote the forward and reverse directions, respectively. (f) The standard deviations of the entity $\mathcal{X}_d$ for both the forward and reverse scenarios are depicted.}
	\end{figure*}
	\begin{figure*}
		\centering 
		\includegraphics[width=\textwidth]{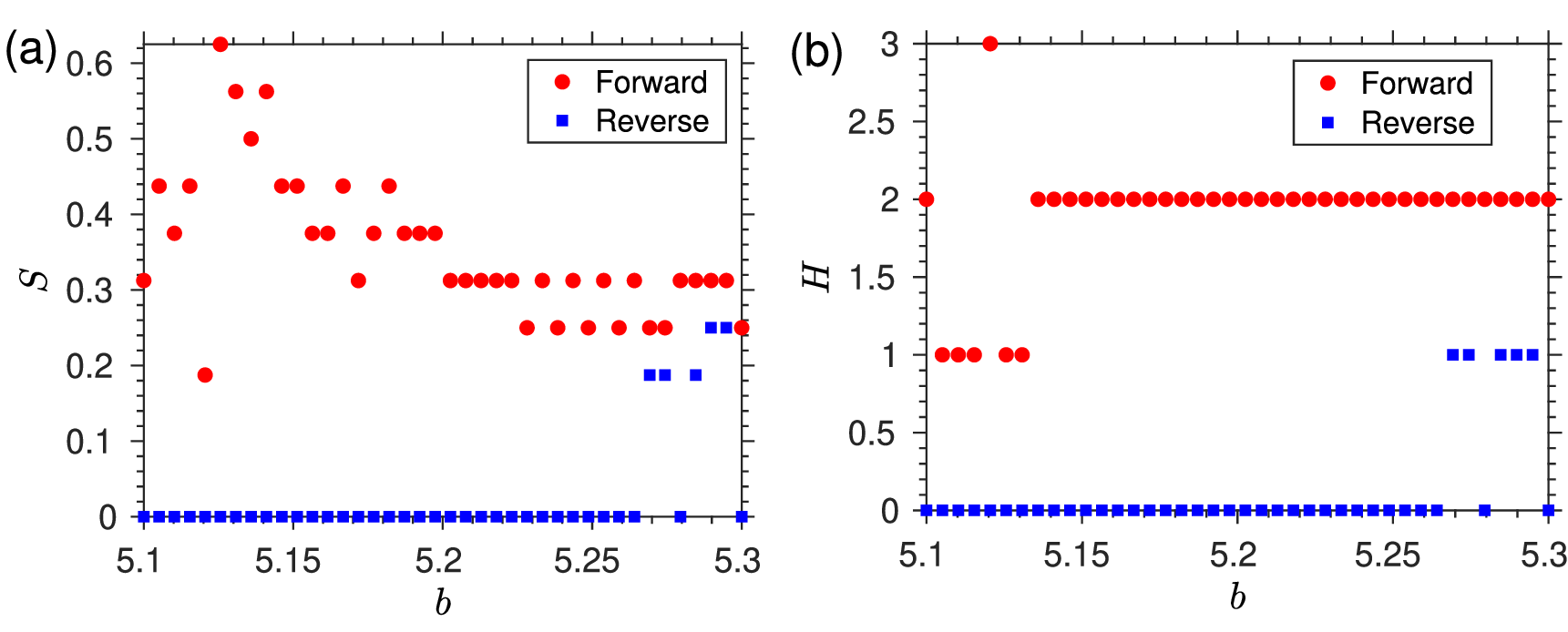}\hfill
		\caption{\label{order}(a) Strength of incoherence (SI), $S$, acquired from the concentration of the activator for the range of control parameter variations in both the forward and reverse directions (b) Corresponding discontinuity measures, $H$}
	\end{figure*}
	\subsection{Protocol-II: reset the dynamics before changing the control parameter}
	In the second protocol, denoted as $\mathcal{RV}$, we also initiate with a uniform state at a particular value of the control parameter, $b_{n=1}$ and generate an amplitude snapshot, $A_{M_1}^\mathcal{RV}$, after the time interval, $\tau$. However, unlike the previous protocol, we enforce a reset mechanism after generating the snapshot at a certain control parameter value. This reset can be visualized as turning off reactions in the chemical system, and $\tau$ is then the time interval between reactions switched on and off. Following this resetting, the system moves to a uniform state corresponding to an upper value of the control parameter, $b_{n=2}$, and time is resorted to $0$. Then, starting from that uniform state, we extract an amplitude trajectory, $A_{M_2}^\mathcal{RV}$, after the same time interval. After the $K$ repetition of this process, we obtain different trajectories corresponding to all the control parameter values at a specific time point ($\tau=2600$). Hence, the spatiotemporal collective concentration dynamics of the Brusselator system in this protocol emerge as,
	\begin{equation}
		{z_{IH_n}}^\mathcal{RV}=z_{I0_n}+A_{M_n}^\mathcal{RV}U_{cH}\exp(if_{cH}\tau)+C.C.,
		\label{hwave}
	\end{equation} 
	with $A_{M_n}^\mathcal{RV}$ being the amplitude field acquired in this protocol at $n$th value of the control parameter. 
	
	In both protocols, the homogeneous steady-state concentration of the system can be represented by $z_{0_n}$ where $z_{I0_n}$ will be the steady-state concentrations of intermediate species corresponding to the $n$th value of the control parameter. The homogeneous steady state of the system depends only on the system parameters. Chemostatted species concentrations remain the same in the inhomogeneous state of the system.
	\section{Nonequilibrium thermodynamic framework \label{ntf}} 
	To capture the thermodynamic entities of the states that appeared in these two protocols, we employ a stochastic thermodynamics-motivated nonequilibrium thermodynamic framework~\cite{Rao2016NonequilibriumThermodynamics, Falasco2018InformationPatterns}. This framework was originally devised for a single CRN. However, the incorporation of effective coupling at the level of the amplitude dynamics facilitates the suitable execution of this nonequilibrium thermodynamic framework for coupled chemical systems~\cite{pkgg3, pkgg6}. We quantify the entropy production rate, work rate, and proper thermodynamic potential of the system in terms of the concentration of the coupled continuum chemical oscillator system, and the concentrations of the intermediate species are in a far from equilibrium state. 

	For the thermodynamic description, we consider two crucial assumptions to be valid. One of these assumptions is the concept of local equilibrium, which allows this nonequilibrium framework to implement the equilibrium form of thermodynamic entities. Further, in the system description, we assume a dilute ideal solution in which chemical species are present as a well-mixed solute, and the solvent in the dilute solution maintains isothermal and isobaric conditions for the system. Hence, we study the nonequilibrium scenario of an isothermal system having multiple chemical reservoirs. We focus on encapsulating the following thermodynamic entities:
	
	\textit{Entropy production rate:} The entropy production rate, one of the central elements in the realm of nonequilibrium thermodynamics, assesses the departure from the detailed balance in the CRN pathway. As the system is operating far from equilibrium, we have a net flux in the system owing to the presence of forces regarding reactions, diffusions, or control parameter variation. We can express entropy production in terms of force and flux (see APPENDIX A). Intermediate species concentrations, $z_{\sigma=I}$, will assume two different values from Eq.~\eqref{chwave} and~\eqref{hwave} depending on protocols. As the total EPR quantifies the entropy change in the system and reservoirs, it is always positive following the second law of thermodynamics.
	
	\textit{Semigrand Gibbs free energy:} Another crucial aspect of nonequilibrium thermodynamics is to construct the proper thermodynamic potential. For this purpose, the role of conservation laws~\cite{Alberty2003ThermodynamicsReactions} of the CRN becomes central. The semigrand Gibbs free energy plays the role of the proper thermodynamic potential of this system~\cite{Falasco2018InformationPatterns} and is denoted as $\mathcal{G}$. For more detail and expression of this semigrand Gibbs free energy, one can consult APPENDIX B.
	
	\textit{Nonconservative work:} In addition to the aforementioned elements, we incorporate the nonconservative work rate, $\dot{w}_{ncon}$ into our nonequilibrium thermodynamics description (see APPENDIX C). Nonconservative work is performed to sustain steady currents of chemostatted species.
	\section{Results and Discussion \label{rnd}}
	\begin{figure*}
		\centering 
		\includegraphics[width=\textwidth]{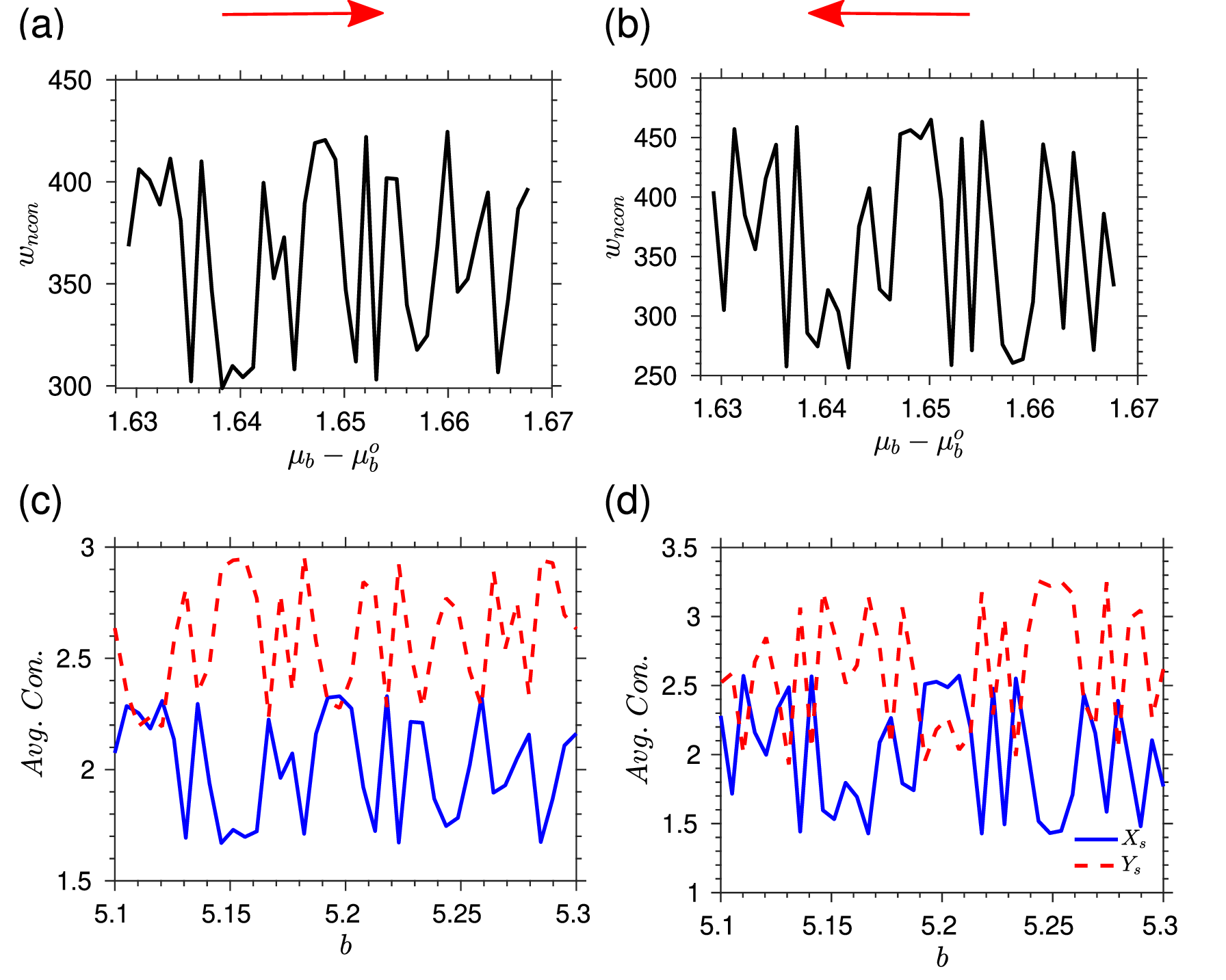}\hfill
		\caption{\label{nonconwork}For protocol-I, nonconservative work profiles with the chemical potential of the control parameter are illustrated in (a) forward, and (b) reverse directions. Arrows refer to the direction of the control parameter variation. For the forward and reverse scenarios, the average concentration dynamics for the control parameter $b$ are displayed in (c) and (d), respectively. The solid blue line represents the concentration of the activator, while the dashed red line portrays the inhibitor concentration. In both directions, nonconservative work profiles qualitatively reflect the average concentration dynamics of the activator.}
	\end{figure*} 
	\begin{figure*}
		\centering 
		\includegraphics[width=\textwidth]{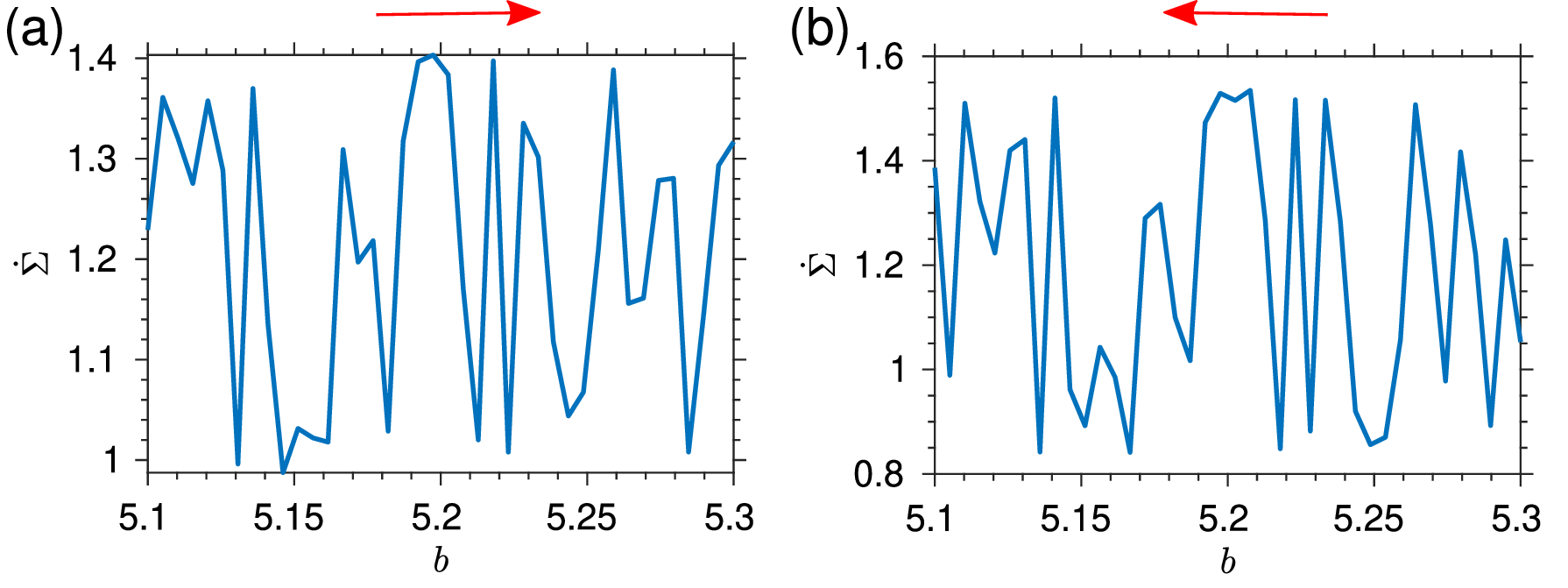}\hfill
		\caption{\label{yaepr}Entropy production rates along the control parameter $b$ for the (a) forward and (b) reverse direction of the protocol. For both cases, entropy production rates exhibit highly irregular oscillatory behavior. Arrows stand for the direction change in the control parameter value.}
	\end{figure*}
	\begin{figure*}
		\centering 
		\includegraphics[width=\textwidth]{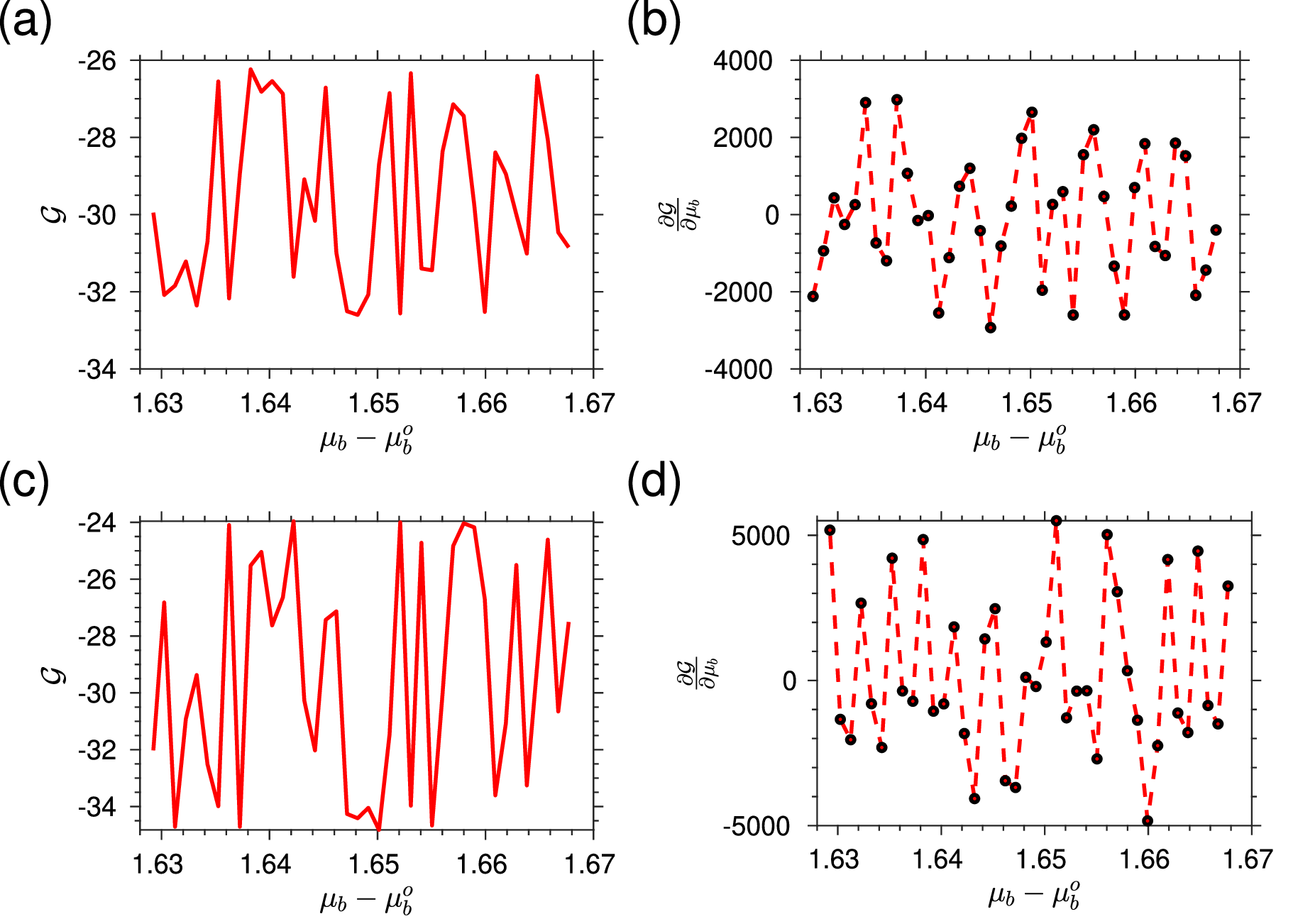}\hfill
		\caption{\label{yaenergy}For the continuous variation of the control parameter, $b$ in the forward direction, (a) the semigrand Gibbs free energy of the inhomogeneous state and (b) its slope concerning the chemical potential of the control parameter. For the same protocol in the reverse direction, (c) and (d) illustrate the semigrand Gibbs free energy and its slope, respectively.}
	\end{figure*}
		\begin{figure*}
		\centering 
		\includegraphics[width=\textwidth]{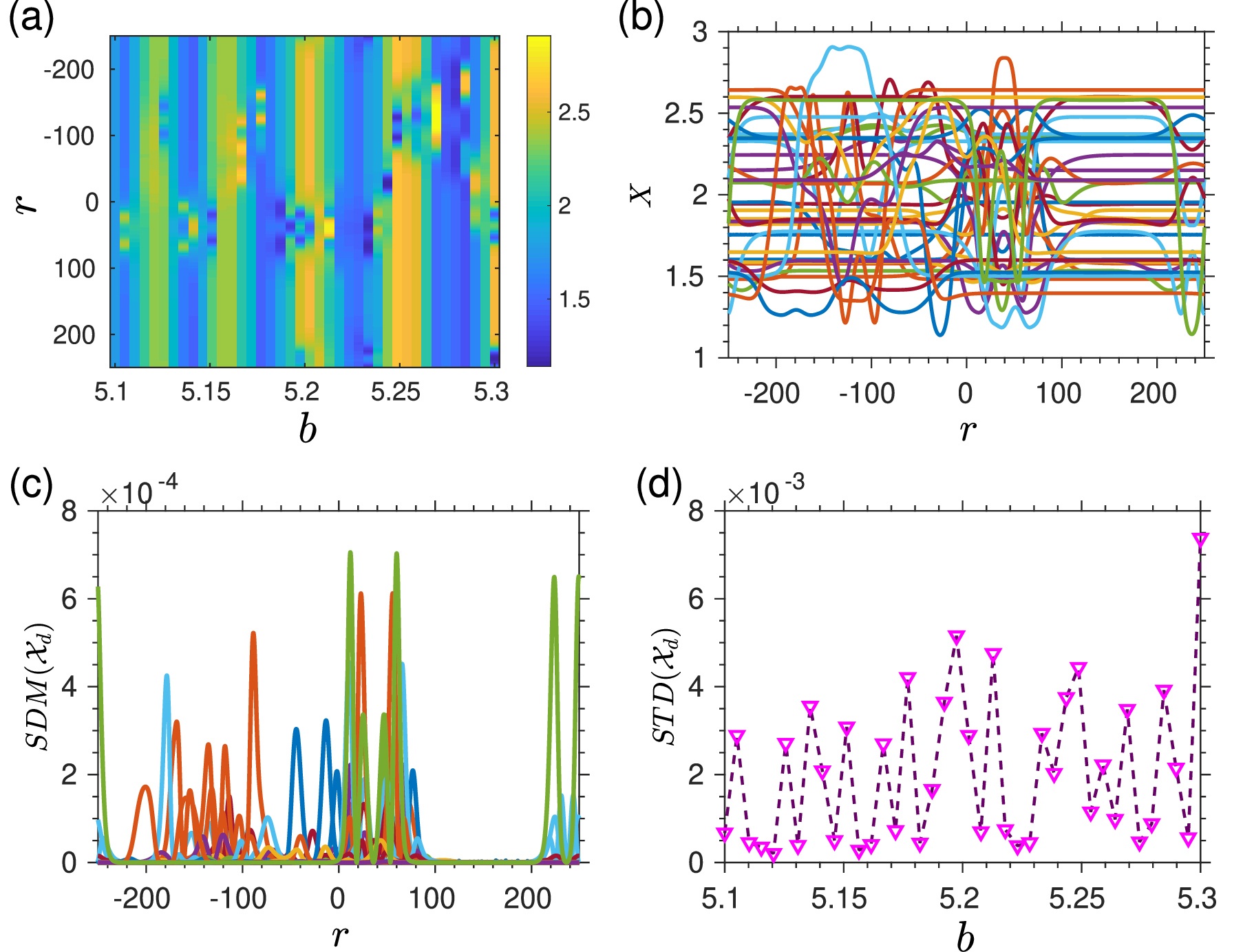}	
		\caption{\label{conreset}(a) Image of the activator concentration field, $X$, obtained using the resetting protocol (b) Line plots of the activator concentration corresponding to different values of the control parameter $b$ over the spatial dimension $r$ (c) Squared deviations from the mean (SDM) of the entity $\mathcal{X}_d$ over the range of control parameter $b$ (d) Standard deviation of the entity $\mathcal{X}_d$ for the resetting protocol.}
	\end{figure*}
	\begin{figure*}
		\centering 
		\includegraphics[width=\textwidth]{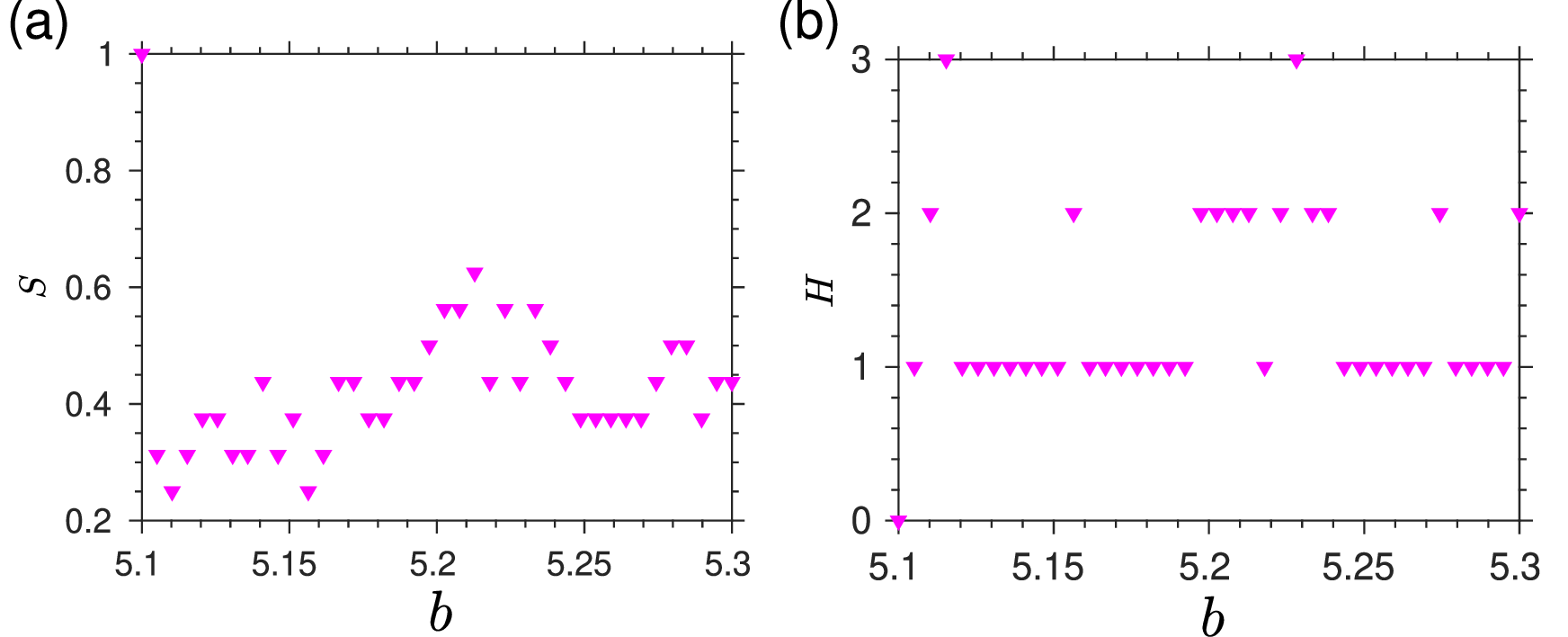}	
		\caption{\label{shrese} In (a), we display the strength of incoherence (SI), S, obtained from the activator concentration for the range of control parameter variations in the resetting protocol. (b) Depicts the corresponding discontinuity measures, H.}
	\end{figure*}    
	\begin{figure*}
		\centering 
		\includegraphics[width=\textwidth]{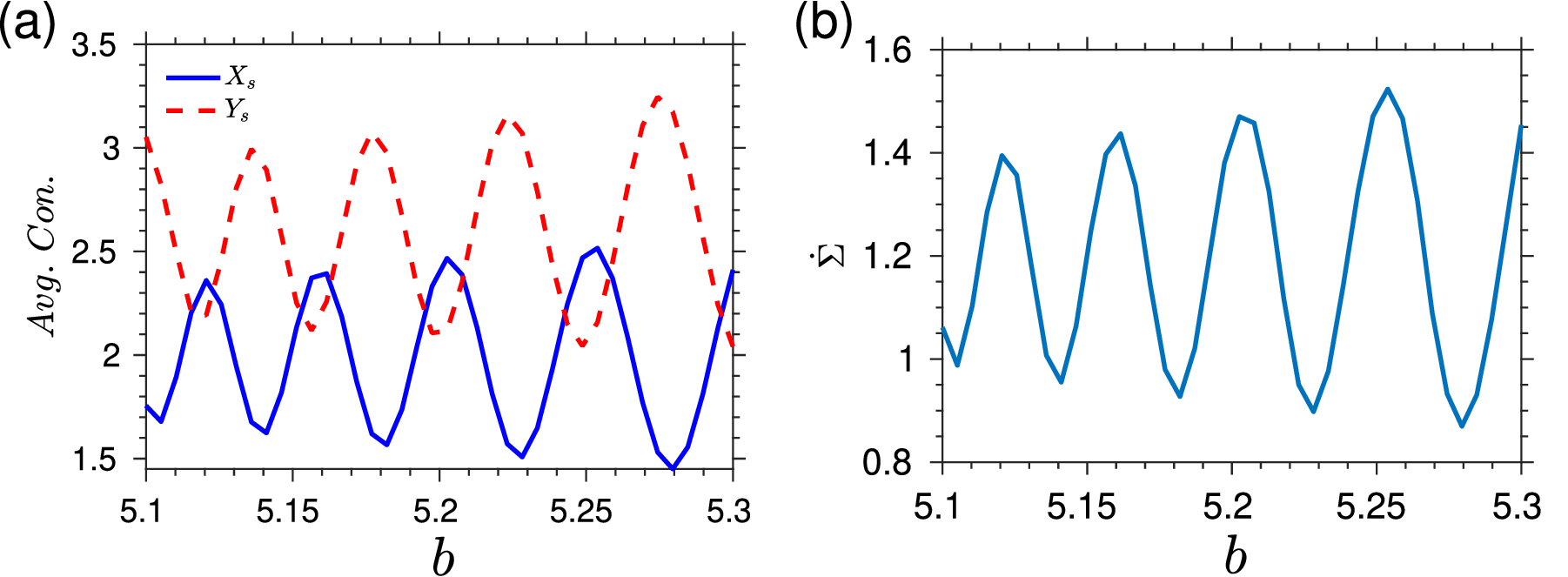}
		\caption{\label{ya1eprreset}(a) The average concentration dynamics for the control parameter $b$, and corresponding (b) Entropy production rate in the resetting protocol. Unlike the previous protocol, both the dynamics and thermodynamics in the resetting protocol demonstrate regular oscillatory behavior over the control parameter $b$.}
	\end{figure*}
	\begin{figure*}
		\centering 
		\includegraphics[width=\textwidth]{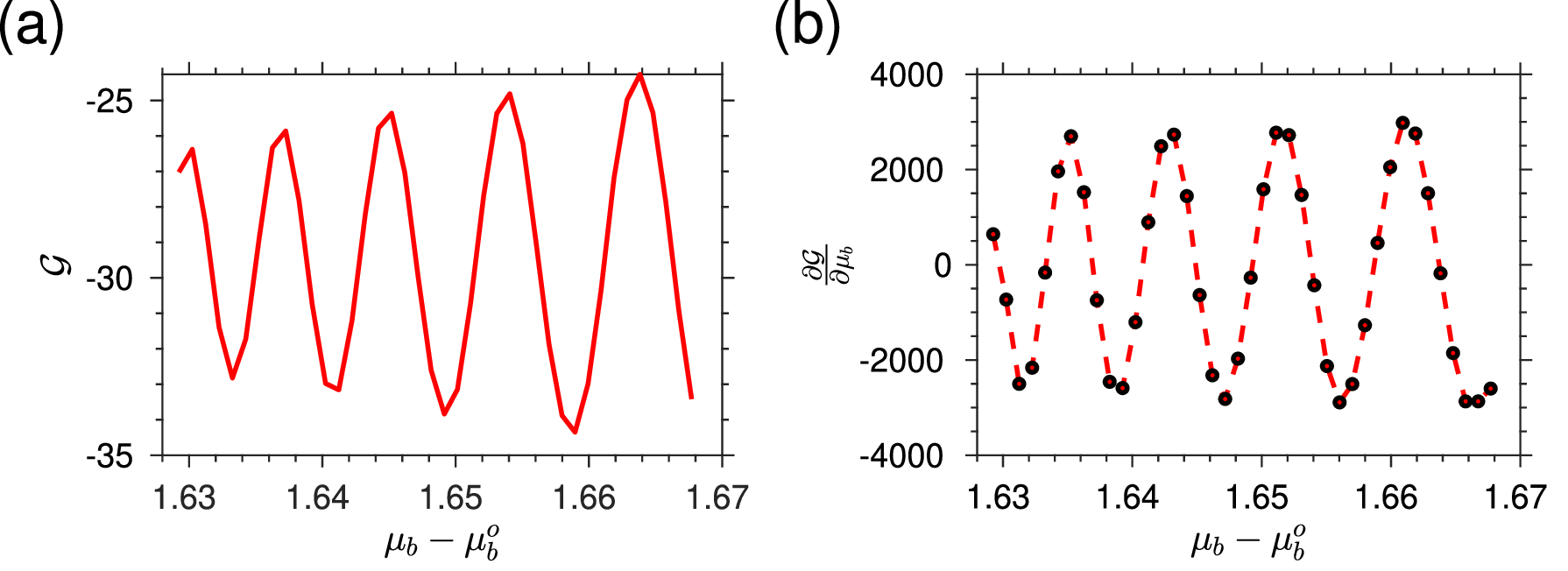}
		\caption{\label{ya1energy}(a) The semigrand Gibbs free energy of the inhomogeneous state and (b) its slope concerning the chemical potential of the control parameter are illustrated for the resetting protocol.}
	\end{figure*}
	All results presented here correspond to an absolute temperature $T= 300 K$, diffusion coefficients 
	$D_{11}=4$, $D_{22}=3.2$, a one-dimensional system of length $l=500$, reverse reaction rate constants $k_{-{\rho}} =10^{-4}$, and forward rate constants, $k_{\rho} = 1$. Steady-state values of two intermediate species in the Brusselator system set the initial uniform base state, ${z_I}_{0}$ in Eq.~\eqref{chwave} and~\eqref{hwave}. In the simulation of the related MCGLE, a time step size of $0.01$ is employed. The one-dimensional system length, $l$, is discretized into 2048 grid points. For convenience, the concentration profiles of the activator in both protocols are denoted simply by $X$ in the following discussions.
	\subsection{Protocol-I results: Memory effect on the dynamics and thermodynamics}  
	The image of the activator concentration field $X$ for the forward variation of the control parameter $b$ in Fig.~\ref{yaconcentration} (b) reveals the presence of a patch-like regime over a specific part of the spatial dimension. For a more apparent visualization of this patch-like regime, line plots of the activator concentration $X$ are depicted in Fig.~\ref{yaconcentration} (a), where each line represents a concentration profile corresponding to a particular control parameter value and is obtained from Eq.~\eqref{chwave} for this protocol. We observe that spatial coherence or incoherence behaviors of concentration profiles corresponding to different control parameter values appear to be locked over a specific spatial part. However, the phases of these profiles can be different.  
	
	To gain a better understanding of this concentration dynamics, we generate a statistical measure called the squared deviations from the mean (SDM) of a newly defined entity $\mathcal{X}_d$. For each control parameter value, a transformed variable $\mathcal{X}_d$ is derived from the activator concentration $X$ as $\mathcal{X}_d=X_{i}-X_{i+1}$, where $i=1,2,..., N-1$ represents the grid point over the spatial length, $l$. This transformed variable, $\mathcal{X}_d$, is defined only at a fixed time point. For the forward control parameter change, the SDMs are denoted by $SDM^f (\mathcal{X}_d)$ and are displayed in Fig.~\ref{yaconcentration} (d) along the spatial dimension. In this figure, the flat regions with vanishingly small values in each profile indicate coherent regimes, while the peaks with relatively higher values imply incoherent parts. From Fig.~\ref{yaconcentration} (d), it is evident that coherent and incoherent regimes of the concentration profiles for most of the control parameter values appear over specific spatial parts, as appeared from the image and line plots of activator concentration.
	
	Next, we employ the same protocol in the reverse direction, starting with the upper end of the control parameter range (i.e., $b=5.30$) and terminating at the lower end of the control parameter (i.e., $b=5.10$). In this reverse counterpart of the protocol, we acquire a different concentration field of activator in Fig.~\ref{yaconcentration} (c). One can notice that the coexistence of coherence and incoherence behavior is only yielded in some initial control parameter values. For all other values, only coherent behaviors emerge. For this reverse counterpart of the protocol, we now derive the same transformed variable $\mathcal{X}_d$ for each control parameter value. The SDM in this reverse counterpart, denoted as $SDM^r(\mathcal{X}_d)$, are then presented along the spatial dimension in Fig.~\ref{yaconcentration} (e). In the reverse direction of the protocol, the SDM measure reveals only a few nearby peaks corresponding to the incoherent behavior of the concentration dynamics. Further on, we obtain the standard deviation of $\mathcal{X}_d$ for both the forward and reverse scenarios in Fig.~\ref{yaconcentration} (f). In this figure, higher values of the standard deviation are related to incoherent or chimera states, while vanishingly small values are associated with the coherent state~\cite{laks}. Hence, from Fig.~\ref{yaconcentration} (f), it is evident that concentration profiles for the forward direction of the protocol have either chimera or incoherent nature for all values of the control parameter. Whereas, in the reverse direction, the concentration profiles predominantly exhibit a coherent nature. The above discussions highlight the contrasting dynamics concerning the control parameter observed in the forward and reverse directions of this protocol.                 

	To further distinguish between chimera and incoherent nature in the concentration profiles, we employ a measure, the local standard deviation $\sigma_{loc}$, calculated using a carefully chosen number of bins $P$. We then use a Heaviside step function $\Theta(\delta-\sigma_{loc})$ with a threshold value $\delta=0.0005|\mathcal{X}_d{\text{max}}-\mathcal{X}_d{\text{min}}|$. Eventually, the strength of incoherence (SI)~\cite{laks} is obtained as $S=1-\frac{\sum_{p=1}^{P}\Theta(\delta-\sigma_{loc}(p))}{P}.$ Another entity, discontinuity measure, $H=\frac{\sum_{i=1}^{P}|\Theta_i-\Theta_{i+1}|}{2}$ is implemented to differentiate between multichimera and chimera states. Figure~\ref{order} displays these qualitative metrics for both the forward and reverse directions of the protocol. By combining the values of $S$ and $H$, we can conveniently identify different states of the concentration profiles~\cite{laks}. Specifically, $(S=0, H=0)$, $(S=1, H=0)$, $(0<S<1, H=1)$, $(0<S<1, 2\leq H\leq\frac{P}{2})$ correspond to coherent, incoherent, chimera, and multichimera states, respectively. In Fig.~\ref{order} (a), for the forward direction, the nonzero and less than $1$ values of $S$ (represented by red circles) indicate that all the concentration profiles correspond to chimera states. Now, taking $H$ values in Fig.~\ref{order} (a) into consideration, we observe that multichimera behavior is predominant within the control parameter range. For the reverse direction, Fig.~\ref{order} (a) shows that for most of the control parameter values, $S=0$, indicating a coherent state. The corresponding values of $H$ in Fig.~\ref{order} (b), are also $0$, fulfilling the criteria, $(S=0, H=0)$ for the coherent state. A few control parameter values generate chimera states, as evident by $(0<S<1, H=1)$ in both figures. Thus, the combined values of $S$ and $H$ aid us in identifying the different states exhibited by the concentration profiles in both the forward and reverse directions.                                                        
	
	In the forward direction of this protocol, the profile of the space-averaged concentration dynamics reveals a highly irregular oscillatory behavior corresponding to different control parameter values, despite the spatial synchronization observed in the concentration profiles for this whole range of the control parameter. This irregular behavior in the space-averaged concentration dynamics is qualitatively analogous to the evolution of nonconservative work presented in Fig.~\ref{nonconwork} (a) and (c) with the chemical potential of the control parameter. Similarly, for the reverse direction of the protocol, the work profile qualitatively reflects the behavior of the averaged concentration as seen from Fig.~\ref{nonconwork} (b) and (d). The highly irregular behavior is also observed in the variation of entropy production rate with the chemostatted species concentration, $b$, for both the forward and reverse scenarios of the protocols in Fig~\ref{yaepr} (a) and (b), respectively. This irregularity is also reflected in the profiles of the semigrand Gibbs free energy and its slope concerning the chemical potential of the reference species $B$. In both scenarios of the protocol, semigrand Gibbs free energy with the chemical potential of $b$ and entropy production rate with the concentration, $b$ exhibit similar profiles, but they are entirely out of phase.
	
	The results above highlight that despite the reverse direction of the protocol predominantly generating coherent states with only a few chimera states, we observe highly irregular behavior in the dynamic and thermodynamic entities for both the forward and reverse directions of the protocol. This feature suggests that the irregular behavior is not related to the dynamic nature of the concentration profiles but is a characteristic of the protocol itself. Hence, regardless of the specific dynamics of the coupled chemical system, the continuous variation protocol of the control parameter in any direction can generate highly irregular changes in the averaged dynamics of the system and also in the energetic and entropic cost of the system.
	\subsection{Protocol-II results: Effect of resetting on the dynamics and thermodynamics} 
	In this part, we explore the concentration dynamics of the system at a fixed time point for the same range of the control parameter with resetting protocol. The image of the concentration field in Fig.~\ref{conreset} (a) illustrates the coexistence of coherent and incoherent behavior throughout the control parameter range. From the line plots in Fig.~\ref{conreset} (b), it can be observed that the coherence and incoherence parts of concentration profiles are scattered over the spatial dimension. All the individual chemical profiles being completely disconnected over the spatial domain is due to the implementation of resetting after each control parameter value. Now we illustrate the squared deviations from the mean, $SDM(\mathcal{X}_d)$, along the spatial length in Fig.~\ref{conreset} (c) by obtaining the same transformed variable $\mathcal{X}_d$ in this resetting protocol. The SDM highlights that most peaks corresponding to the incoherent part of the concentration profiles are scattered over the spatial length. However, within a small regime on the spatial dimension (between $r=100$ and $200$), we obtain vanishingly small values, indicating coherence. In Fig.~\ref{conreset} (c), the line plots further demonstrate that within the above-mentioned region, the amplitude of the concentration profiles for different control parameter values exhibit similar characteristics. This observation explains the exceptionally small values of $SDM(\mathcal{X}_d)$ in that particular regime. The finite standard deviation of the entity $\mathcal{X}_d$ in Fig.~\ref{conreset} (d) implies that the concentration profiles associated with all values of the control parameter correspond to either incoherent or chimera states in this protocol. To distinguish between incoherence and chimera profiles, the strength of incoherence, $S$, is shown in Fig.~\ref{shrese} (a). All $S$ values are nonzero and less than $1$ (represented by a downward-pointing triangle) except for the initial control parameter value where $S=1$ indicates an incoherent state. For all other concentration profiles, the existence of chimera states is designated. The discontinuity measure, $H$, in Fig.~\ref{shrese} (b) suggests the presence of classical chimera states for many control parameter values. However, within the middle range of the control parameter, a predominant multichimera behavior is observed.    
	
	In this resetting protocol, the space-averaged concentration demonstrates almost regular oscillatory behavior for the whole range of the control parameter in Fig.~\ref{ya1eprreset} (a). Similar to the previous protocol, this nonconservative work (not shown) reflects the same qualitative behavior as the space-averaged concentration dynamics. We also obtain the entropy production rate incurred throughout the process in Fig.~\ref{ya1eprreset} (b). This entropy production rate also exhibits oscillatory change with the control parameter. Previously, this qualitative similarity between entropy production and averaged concentration dynamics has been reported for individual chemical and biological oscillators having spatiotemporal dynamics~\cite{pkgg,pkgg2}.     
	
	Similar to the entropy production rate, the semigrand Gibbs free energy profile in Fig.~\ref{ya1energy} (a) also has oscillatory behavior concerning the chemical potential of the control parameter. However, entropy production rate and semigrand Gibbs free energy profiles are out of phase. The corresponding slope is presented in Fig.~\ref{ya1energy} (b). The scattered spatial dynamics in this protocol concerning the appearance of the incoherence state yield a smooth and almost regular oscillatory behavior in dynamic and thermodynamic space with the variation of the control parameter.          
	\section{Conclusion \label{conu}}
	In this investigation, for two different protocols, we obtain completely different space-averaged profiles of concentration, entropy production rate, and semigrand Gibbs free energy within the same chemostatted environment, which illustrates the sensitivity of the coupled dynamics and thermodynamics to the way of control parameter change. From discussions related to the variation of dynamic and thermodynamic entities in both protocols, it is possible to recognize those differences qualitatively. While the resetting protocol brings randomness in concentration dynamics corresponding to the variable chemostatted environment, the continuity scheme generates a patch-like feature in the system for the forward direction. Moreover, in the forward direction of protocol-I, multichimera behavior dominates, whereas classical chimera profiles appear most frequently in the resetting protocol. While in the resetting protocol, the variation of the dynamic and thermodynamic entities with the control parameter shows oscillatory behavior, the same entities for the continuous variation of the control parameter exhibit highly irregular profiles.
	
	Despite employing two protocols within the specific framework of a chemical oscillator, our findings are applicable to a variety of systems where similar coupled dynamics are observed. This indicates that implementing these protocols could be viable in diverse setups. These comparative studies and insights related to the two different protocols of control parameter change can aid in desirably controlling such coupled chemical systems. This study holds the potential to aid in reservoir computing~\cite{tanaka2019recent} study using a coupled chemical oscillatory system as a reservoir. This study in continuum settings could be performed in a ring of coupled oscillators, thereby enabling similar protocols in Kuramoto-type networks~\cite{Kotwal}. Additionally, the distinct results emerged in two protocols here for the globally coupled oscillator can be extended to the local and nonlocal coupling schemes to assess the generality of these outcomes and protocols. One intriguing direction for future research could involve numerically investigating similar protocols by directly incorporating coupling into the Brusselator RDS, operating beyond the Hopf instability, and formulating a thermodynamically consistent description to capture central nonequilibrium thermodynamic entities of such coupled RDS.  
	
	We have not included the cost of external driving needed for the proper implementation of these protocols in this current study. Including those costs may calculate the thermodynamic costs more accurately. However, those costs will depend on the exact mechanism employed for achieving those protocols. More concrete studies regarding this point will be addressed in future work. Further, by choosing the range of the control parameter, this work constrains itself to some particular states among various possible collective states in coupled chemical oscillator systems. 
	\section*{\label{NTF}APPENDIX A: Expressions of entropy production rates}
	The reaction flux due to the thermodynamic driving force of a reaction, reaction affinity~\cite{Prigogine1954ChemicalDefay.}, follows the mass action law. The forward or reverse flux of an elementary reaction reads, $j_{\pm \rho}=k_{\pm\rho}\prod_{\sigma}z^{v_{\pm\rho}^{\sigma}}_{\sigma}$ with $z_{\sigma}$ being the concentration of the $`\sigma'$ species. Here, $'+'$ and $'-'$ imply the forward and backward reactions, respectively, and $v_{\pm \rho}^{\sigma}$ denotes the number of molecules of a particular species, $`\sigma'$. The net flux is then $j_{\rho}=j_{+\rho}-j_{-\rho}$. The reaction affinity, $f_{\rho}$, can be expressed as $f_{\rho}=-\sum_{\sigma}{S_{\rho}^{\sigma}\mu_{\sigma}}$, where $S_{\rho}^{\sigma}=v_{{-}\rho}^{\sigma}-v_{{+}\rho}^{\sigma}$ is the stoichiometric coefficient of species and $\mu_{\sigma}=\mu_{\sigma}^o+\ln{\frac{z_{\sigma}}{z_0}}$ renders the nonequilibrium chemical potential. In chemical potential expression, $z_0$ represents the solvent concentration, and $\mu_{\sigma}^o$ denotes the standard-state chemical potential. The local detailed balance condition and flux-force form facilitate the expression of the entropy production rate (EPR) of the reaction pathway as $\frac{d\Sigma_{R}}{dt}=\frac{1}{T}\int dr \sum_{\rho} (j_{+\rho}-j_{-\rho}) \ln{\frac{j_{+\rho}}{j_{-\rho}}}.$ Here, the solvent set the constant absolute temperature $T$, and for convenience, $RT$ is taken to be unity. Similar to the reaction part, using diffusive flux and affinity, the entropy production rate due to diffusion becomes $\frac{d\Sigma_{D}}{dt}=\int dr \Big[ D_{11}\frac{{\parallel{\frac{\partial x}{\partial r} }\parallel}^2}{x}+D_{22}\frac{{\parallel{\frac{\partial y}{\partial r} }\parallel}^2}{y}\Big]$. The total entropy production rate includes both the reaction and the diffusion entropy production rates. 
	\section*{APPENDIX B: Conservation laws and Semigrand Gibbs free energy} Conservation laws are the left null eigenvectors of the stoichiometric matrix $S_{\rho}^{\sigma}$ of the CRN, 
	$\sum_{\sigma}{l_{\sigma}^{\lambda}S_{\rho}^{\sigma}}=0$, where $\{l_{\sigma}^{\lambda}\}\in \mathbb{R}^{(\sigma-w )\times \sigma}, w=rank(S_{\rho}^{\sigma}).$ These conservation laws renders components, $L_{\lambda}=\sum_{\sigma}{l_{\sigma}^{\lambda}}z_{\sigma}$. Components are globally conserved quantities of a closed system. Bruseelator CRN possesses two components, $L_1=x+y+a+e$ and $L_2=b+d$. When the Brusselator reaction network is opened by chemostating, two conservation laws break, and corresponding components no longer remain as globally conserved quantities. Based on participating in breaking the conservation laws, we can divide the set of chemostatted species into two subsets in the open system $\{C\}=\{C_{b}\}\cup\{C_{u}\}$. Here, the pair of chemostatted species, $A$ and $B$, are considered to break both conservation laws of the Brusselator network and belong to the subset $C_{b}$. Hence, $A$ and $B$ serve as reference chemostatted species. 
	
	Nonequilibrium Gibbs free energy~\cite{Fermi1956Thermodynamics} of the reaction network can be expressed as
	$G=G_{0}+ \sum_{\sigma \neq 0}{(z_{\sigma}\mu_{\sigma}-z_{\sigma})}\label{neg},$ with constant $G_{0}$ being the solvent contribution. However, due to broken conservation laws, nonequilibrium Gibbs free energy no longer captures the proper energetics of the system. Instead, a transformed Gibbs free energy~\cite{Rao2016NonequilibriumThermodynamics}, the semigrand Gibbs free energy~\cite{Falasco2018InformationPatterns}, $\mathcal{G}$, properly quantifies the energetics of the system. The semigrand Gibbs free energy is obtained by subtracting the energetic cost of exchanged moieties from the nonequilibrium Gibbs free energy,
	$\mathcal{G}=G-\sum_{C_b}{\mu_{C_b}M_{C_b}}$, where $M_{C_b}$ and $\mu_{C_b}$ are the concentration of moieties and reference chemical potential, respectively.
	\section*{APPENDIX C: Nonconservative work rate} 
	In our system, chemostatted species of CRN are homogeneously distributed, and hence reference chemical potential is equal to the chemical potential of the reference chemostatted species, $\mu_{C_b}^{ref}=\mu_{C_b}$. This reference chemical potential leads to the vanishing of the fundamental forces~\cite{chemicalwaveespasito} associated with the reference chemostatted species, i.e., $\mathcal{F}_{C_b}=0$. Specifically, for the Brusselator system, the homogeneous distribution of chemostatted species results in $\mu_{a}^{ref}=\mu_{a}$ and $\mu_{b}^{ref}=\mu_{b}$. Whereas, for the chemostatted species of the subset $C_{u}$, fundamental force can be expressed as $\mathcal{F}_{C_u}=\mu_{C_u}-\sum_{C_b}\mu_{C_b}\sum_{\lambda_b}{l_{C_b}^{\lambda_b}}^{-1}l_{C_u}^{\lambda_b}$ taking into account the difference in chemical potentials of dissimilar chemostats. Consequently, for the Brusselator system, we write $\mathcal{F}_{d}=\mu_{d}-\mu_{b}$ and $\mathcal{F}_{e}=\mu_{e}-\mu_{a}$, implying $\mu_{d}^{ref}=\mu_{b}$, and $\mu_{e}^{ref}=\mu_{a}$. Finally, the nonconservative work in such a system can be defined as $\dot{w}_{ncon}=-\sum_{C_u}(\mu_{C_u}-\mu_{C_u}^{ref})S_{\rho}^{C_u}j_{\rho}.$ This nonconservative work keeps the concentration of species in the $C_u$ subset constant by counteracting the effect of the chemical reactions.
	\bibliography{references.bib}
\end{document}